\documentclass[%
reprint,
 superscriptaddress,
nofootinbib,
 amsmath,amssymb,
pra,
floatfix,
]{revtex4-2}

\usepackage{mathtools}

\usepackage{lipsum} 
\usepackage[position=top]{subfig}
\usepackage{graphicx}
\usepackage{dcolumn}
\usepackage{bm}
\usepackage[hidelinks,colorlinks=true,linkcolor=blue,citecolor=blue]{hyperref}

\usepackage{dsfont}
\usepackage{subcaption}
\usepackage{graphicx}
\usepackage{adjustbox}

\usepackage{braket}

\begin{document}

\preprint{APS/123-QED}

\title{Exponential Expansion of Massive Schr\"{o}dinger Cats for Sensing and Entanglement}

\author{Lorenzo Braccini} \email{lorenzo.braccini.18@ucl.ac.uk}
\affiliation{Department of Physics and Astronomy, University College London, Gower Street, WC1E 6BT London, United Kingdom}

\author{Alessio Serafini}
\affiliation{Department of Physics and Astronomy, University College London, Gower Street, WC1E 6BT London, United Kingdom}


\author{Sougato Bose}
\affiliation{Department of Physics and Astronomy, University College London, Gower Street, WC1E 6BT London, United Kingdom}


\begin{abstract}

Schr\"{o}dinger cat states of levitated masses have several applications in sensing and, offer an avenue to explore the fundamental nature -- classical vs nonclassical -- of gravity, eg, through gravitationally induced entanglement (GIE). The interaction between a qubit and a levitated mass is a convenient method to create such a cat state. The size of the superpositions is limited by weak mass-qubit interactions. To overcome this limitation, we propose a protocol that exponentially expands an initially small superposition via Gaussian dynamics and successfully recombines it to complete an interferometry. 
An unknown force can be sensed by the superposition exponentially fast in the expansion time. The entanglement between two such interferometers interacting via a quantum force is -- for the first time in qubit-based non-Gaussian protocols -- obtained by solving the full quantum dynamics using Gaussian techniques. GIE grows exponentially, thereby making it closer to experimental feasibility. Requirements of experimental precision and decoherence are obtained.

\end{abstract}

\maketitle


\textit{Introduction} -- Levitated masses in the quantum regime have applications in testing collapse models, detecting particles, force sensing and studying gravity~\cite{bassi_models_2013, giovannetti_advances_2011,hempston_force_2017,tao_yectonewton_2023, fuchs_measuring_2024, bose_massive_2023}, with one of the most sensitive instance \cite{qvarfort_gravimetry_2018, marshman2020mesoscopic,eva_requirements_2023,barker2022entanglement,toros_relative_2021,wu2023quantum} being large quantum superpositions -- a mass in two separated places at once. Substantial advances have been made in quantum control in this field, including the experimental demonstrations of ground-state cooling~\cite{ballestero_levitodynamics_2021, delic_cooling_2020, tebbenjohanns_motional_2020,magri_real_2021}. Notably, highly delocalized quantum states of two masses can pose fundamental questions on the quantum nature of the gravitational field, via detection of Gravitationally Induced Entanglement (GIE)~\cite{bose2016matter,bose_spin_2017, marletto_gravitationally-induced_2017}. The creation of non-classical states in levitated systems -- more sensitive to forces and GIE -- is an active field of research~\cite{bose_preparation_1997, bose_scheme_1999,Romero-Isart_large_2011, Romero_Isart_quantum_2011, scala2013matter, geraci_sensing_2015}. 

Within non-Gaussian dynamics, the entangling interaction between a levitated mass and external degrees of freedom can create superpositions in the mass position, \text{i.e.}, generate a cat state~\cite{bose_qubit_2006}. Such behavior is predicted to happen with levitated masses with a dopant spin (such as Nitrogen-Vacancy centers)~\cite{scala2013matter, yin_large_2013, wan_free_2016, marshman_constructing_2022}, in magnetic traps with superconducting flux qubits~\cite{nair_massive_2023}, in Paul traps with an atom coupled to an optical field~\cite{toros_creating_2021}, and optomechanical systems~\cite{bose_preparation_1997, bose_scheme_1999,qvarfort_gravimetry_2018}. Other methods, using non-linear potentials are also present in the literature~\cite{ralph_dynamical_2018, roda-llordes_macroscopic_2024}. Within Gaussian evolution, delocalization can be achieved with the spread of the wavefunctions, through free-evolution or squeezing~\cite{giovannetti_quantum_2004, wu_squeezing_2024, miao_quantum_2020}. Two recent works investigate the exponential squeezing generated by inverted quadratic potentials and its application in GIE experiments~\cite{weiss_large_2021, fujita_inverted_2023}. This work exploits elements of this latter approach and shows that it is possible to exponentially separate the components of a small cat state -- a manifestly non-Gaussian state -- purely by Gaussian dynamics. Furthermore, entanglement (including GIE) is -- for the first time -- exactly treated as a fully quantum evolution of the joint non-Gaussian wavefunction of the interacting interferometers, without approximating the wavepackets as semi-classical paths (leading to only phases) or ignoring the creation-recombination processes (for static treatment see Ref. \cite{bengyat_gravity_2023}).

Given qubit state-dependent force $F_q \sigma_z$, the reachable superposition size of the cat state is $\sim F_q/M\omega^2$, where $M$ is the mass and $\omega$ is the trap frequency \cite{scala2013matter}. Due to experimental challenges and fundamental constraints, $F_q$ is usually small in value, limiting the superposition sizes. For example, $F_q$ depends on the Bohr-magneton,  the flux quantum, or the atomic dipole, for dopant spins, superconducting qubits, and optically coupled atoms, respectively. Specifically, for dopant spins, $F_q$ can be increased with a higher magnetic gradient which actually severely limits the superposition size increasing $\omega$~\cite{pedernales_motional_2020,marshman_constructing_2022}. 

Hence, GIE experiments find a serious impediment that limits their implementation. Given a $F_q$, the superposition size can be increased by performing low frequency experiments, so that the GIE can be sensed. However, the required timescales are either higher than the decoherence times of the ancillary qubit or -- assuming long coherence -- approach frequencies lower than $\mathcal{O}(1)$Hz, where the seismic noise is difficult to filter~\cite{abbott_guide_2020}. Different but still limiting problems arise for squeezing protocols in optical traps, where the experimental time is limited by photon recoil~\cite{hoang_electron_2016}. Although possible expansions of the superpositions exploiting initial velocity differences of $\sim F_q/M\omega$ (which are themselves too small), they only grow linearly in time $\tau$ and thereby do not suffice experimental time much greater than $1$s~\cite{pedernales_motional_2020, marshman_constructing_2022,  zhou_catapulting_2022, zhou_mass_2023, Zhou_gravito_2024}. We note that the GIE is not the rate limiting step here, rather it is the creation of a large enough superposition. A quantum dynamics that expands the superposition while significantly reducing the time is missing in the literature. The presented protocol is kept general with the aim of its possible application in different physical systems. 

\textit{Creation of Massive Cat States} --  The mechanism to create cat states in levitated objects has been extensively studied, but we shall review it, to introduce notation and quantities -- see Supplementary Material (SM) for more details and Ref.~\cite{serafini_quantum_2017}. The position and the momentum of a trapped quantum object are described by the observable $\hat{X}$ and $\hat{P}$, with commutation relation $[ \hat{X}, \hat{P}] = i \hbar$, where $\hbar$ is the Plank constant. The ground state spread is given by $x_0 = \sqrt{\hbar/(2 M \omega)}$, where $M$ is the mass and $\omega$ is the frequency of the trap. The unit-less position and momentum operators are $\hat{x} = \hat{X}/(\sqrt{2} x_0)$ and $ \hat{p} = \hat{P} (\sqrt{2} 
x_0/\hbar) $. By defining the vector of operators $\hat{r} := (\hat{x}, \;\; \hat{p})^T $, the commutation relations can be written in compact form as $[\hat{r},\hat{r}^T] = i \Omega$, where $\Omega$ is the 2-dimensional symplectic form. The mass is trapped in a Quantum Harmonic Oscillator (QHO), and the Hamiltonian of the system can be written as $\hat{H} = \hbar \omega (\hat{x}^2 + \hat{p}^2)/2$. Another quantum system -- taken to be a qubit, and generalizable to larger spins~\cite{braccini_large_2023, steiner_pentacene_2024} -- can employ an operator-valued force $F_q$ to the mass. This can be described by the interaction Hamiltonian $ - F_q x_0 \hat{x} \otimes \hat{\sigma}_z $, where $\hat{\sigma}_z $ is the unit-less Pauli operator.  By preparing, the qubit in a superposition state $\frac{1}{\sqrt{2}}(\ket{+1}_q+\ket{-1}_q)$ and the mass in the ground state $\ket{0}_m$, the interaction leads to the creation of a cat state. The time is taken to be in  units scaled by the frequency, i.e. $\omega t \to  t$ so that it is dimensionless. At time $t_0 = \pi$, the state -- up to trivial dynamical phases of the qubit -- is
\begin{equation}
\label{eq:cat_state}
    \ket{\psi(0)} = \frac{1}{\sqrt{2}}\left(\ket{r_0 }_m \ket{+1}_q + \ket{-r_0}_m \ket{-1}_q \right),
\end{equation}
where $\ket{r} := \hat{\mathcal{D}}_r \ket{0}$ is a Coherent State (CS), $\hat{\mathcal{D}}_r = e^{i r^T \Omega \hat{r}}$ is the displacement operator, and $r_0 = \left(\delta x,    \;\;\;\; 0 \right)^T$  where $\delta x = 2 F_q x_0 / ( \hbar \omega)$, is half of the maximum spatial initial superposition size, in ground-spread units. As previously claimed, in dimension-full units, the superposition is $x_0 \delta x = F_q/M \omega^2$. In the usual treatment, the mass is let evolve under the same dynamics for another time $t_0 = \pi$ such that one-loop interferometry can be completed and the mass position recombined~\cite{scala2013matter}. Instead, we assume that the interaction Hamiltonian can be switched off -- as discussed more in detail in the conclusion -- and the mass undergoes the protocol presented below. $\delta x$ is kept as a free parameter.  

\begin{figure}[!t]
\includegraphics[width=0.5\textwidth]{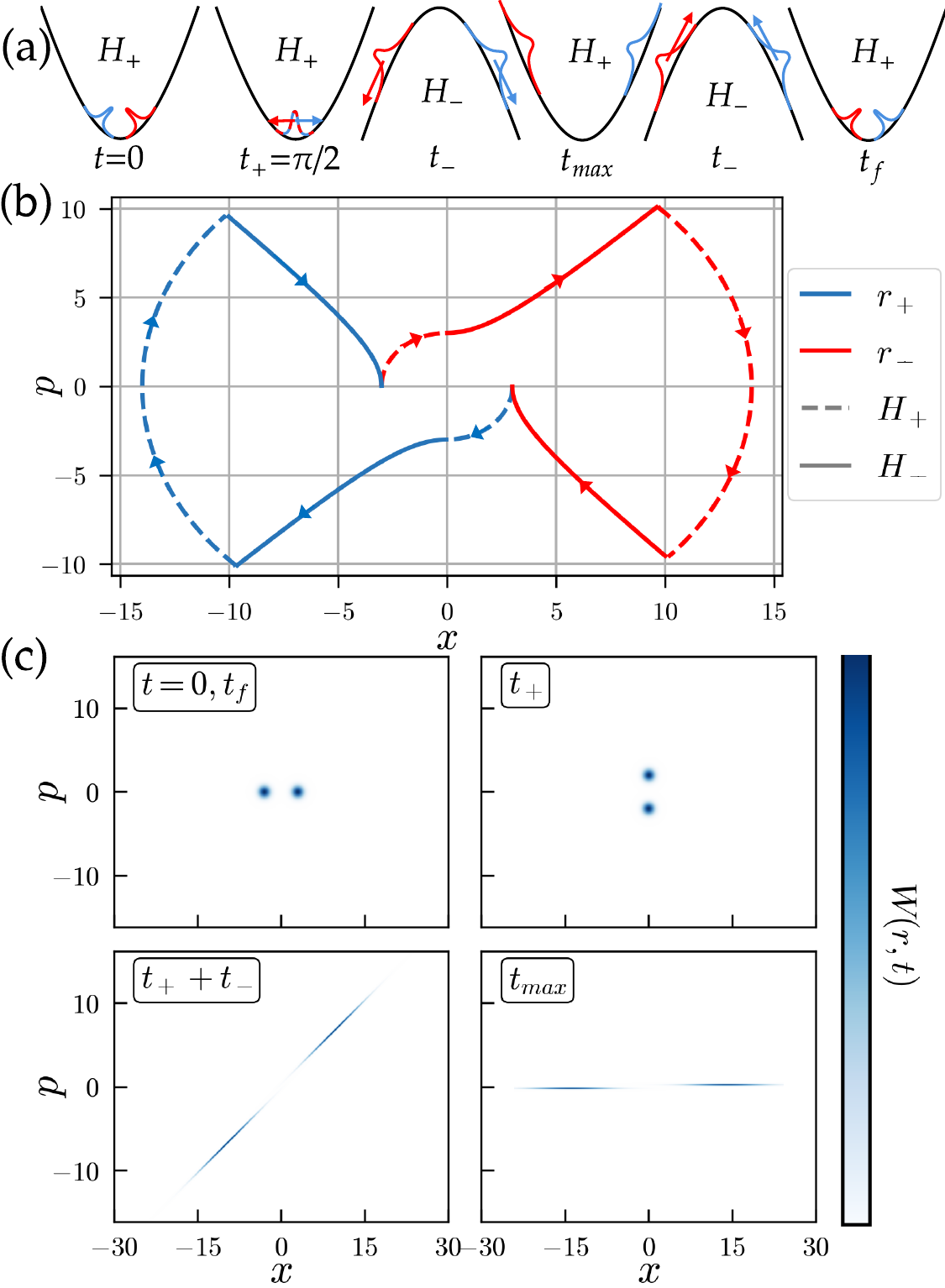}
  \caption{\label{fig:first} Representation of a cat state evolving in the expansion dynamics: (a) Schematic representation of the expansion protocol; (b) Time evolution of the first moments associated with the $\ket{\pm 1}$ qubit states (initial superposition $\delta x = 3$ and expansion time $t_- = 0.6 \pi$); (c) Wigner functions of the reduced density matrix of the position degrees of freedom of the levitated mass at different times of the protocol -- showing that each component remains Gaussian, albeit squeezed.}
\end{figure}

\textit{Expansion and Completion of Interferometry} --  The protocol consists of an expansion and recombination process, performed one after the other, starting at $t_0=\pi$, which is taken to be $t=0$ for the protocol.  This choice of $t_0$ is found to be optimal for the following discussions -- a different $t_0$ would have led to smaller position superposition but larger momentum ones. Both expansion and recombination processes are made of one evolution in a QHO for a time $t_+=\pi/2$, followed by an evolution for a time $t_-$ in an IHO, with Hamiltonian $H_- = \hbar \omega (\hat{p}^2- \hat{x}^2 )/2$.  A representation of the protocol is given in Fig.~\ref{fig:first}. For the following treatment, the two potentials are taken to have the same \textit{frequency} $\omega$~\footnote{In experimental settings, it may not be possible to create high-frequency IHO. The dynamics of the expansion will be scaled by a factor of $\omega/\omega_I$, where $\omega_I$ is the frequency of the IHO.}. In this paragraph, for simplicity of expressions, the hamiltonian will be scaled according $\hat{H} \to \hat{H}/(\hbar \omega)$. The four switching times are assumed to be fast enough such that the sudden approximation holds. Then, the total unitary at $t_f = 2 \left( t_+ + t_-\right)$ is given by
\begin{equation}
\label{eq:unitary_ideal}
    \hat{S}_{\text{tot}} =e^{-i  t_- \hat{H}_-} e^{-i t_+ \hat{H}_+}   e^{-i  t_- \hat{H}_-} e^{-i t_+ \hat{H}_+} \;.
\end{equation}
While the studied states are highly non-Gaussian, both the dynamics and the wave-functions entangled with each $\ket{\pm 1}_q $ qubit state are Gaussian. Hence, the two superposed states are fully characterized by their first and second moments. For a general Gaussian state $\rho(t)$, they are defined according to $r(t) := \text{Tr} [ \rho(t) \hat{r}] $ and
\begin{equation}
    \sigma(t) := \text{Tr} [ \{ (\hat{r}-r), (\hat{r}-r)^T \} \rho(t)] \;,
\end{equation}
where $ \{ \cdot, \cdot \}$ denotes anti-commutation. The Hamiltonians are quadratic in $\hat{x}$ and $\hat{p}$ and, hence, there is no scope outside of a Gaussian dynamics. Thus, it can be described as a symplectic transformation of the phase space (see SM). Let us denote $S(t)^{-1}$ as the symplectic transformation up to time $t$. Then, the time evolution of the first and second moments are $r_\pm(t) = \pm S(t)^{-1} r_0 $ and $\sigma_\pm(t) = S(t)^{-1} \sigma_\pm(0) \left(S(t)^{-1}\right)^T$, where $\pm$ refers to the state entangled with $\ket{\pm1}_q$, and $\sigma_\pm(0) = \mathds{1} $ for initial cat states of CS (Eq.~\ref{eq:cat_state}). The time evolution of the first moments and the Wigner function of the total state are plotted in Fig.~\ref{fig:first}. At time $t_\text{max} = t_- + 3 \pi/4 $, the maximum superposition is achieved, of value $\Delta x = \sqrt{2} e^{t-} \delta x $, \textit{i.e.} exponentially expanding the superposition in $t_-$.

 At the end of the expansion protocol, $t_f = 2 (t_+ + t_{-})$ and the total symplectic transformation (to not be confused with the unitary $\hat{S}_{\text{tot}}$) is 
\begin{equation}
\label{eq:symplectic_ideal}
    S_{\text{tot}}^{-1} = S_{H_-}^{-1} S_{H_+}^{-1} S_{H_-}^{-1} S_{H_+}^{-1} = - \mathds{1} \;,
\end{equation}
where $S_{H_\pm} = e^{- t_\pm \Omega H_\pm }$, and the $H_\pm$ matrix is such that $\hat{H}_\pm = \hat{r}^T H_\pm \hat{r} / 2$. The second identity follows by noticing that $S_{H_+} = - \Omega$ (for $t_+=\pi/2$), $S_{H_-} = S_{H_-}^T$ (for all $t_-$), and by applying the definition of symplectic group ($S \Omega S^T = \Omega$) and $\Omega^2 = - \mathds{1}$. Hence, $r_\pm(t_f) = \mp r_0 $ and  $\sigma_\pm(t_f) = \sigma(0)$, \textit{i.e.} the CSs are mapped back to their original superposition but swapped, equivalent to having performed a spin flip on the qubit. Hence, by applying a $\pi$-pulse on the qubit, the state is given by Eq.~\ref{eq:cat_state}. By letting the mass-qubit system interact again and evolve for $t_0 = \pi$, the mass can be recombined -- as in original one-loop protocols. The final state of the system is $\frac{1}{\sqrt{2}} \ket{0}_m (\ket{+1}_q+\ket{-1}_q)$. The mass degree of freedom -- being in a product state -- can be trivially traced out, implying no loss of visibility in the measurements on the qubit. The interferometry is then completed. 

\textit{Force Sensing} -- In the following section, an unknown force $f$ is applied to the expanding cat state. The force introduces a linear term in the Hamiltonian of the form $\hat{H}_f = - f \hat{X}$, so that $\hat{H}_\pm^f = \hat{H}_\pm +  \hat{H}_f $. Up to an overall phase, any general Gaussian dynamics -- such as the proposed protocol with the external force -- can be expressed as $\hat{U} =  \hat{\mathcal{D}}_{r_\text{tot}} \hat{S}_{\text{tot}} $, where  $\hat{\mathcal{D}}_{r_\text{tot}}$ is the total displacement operator and $\hat{S}_{\text{tot}} $ the total squeezing (see SM). At the end of the interferometer, the reduced qubit density matrix ($\rho_q$) can be computed by tracing out the position degree of freedom. The final qubit state is $\ket{\psi}_q = \frac{1}{2} (\ket{0}_q + e^{ i \phi_f} \ket{1}_q )$ where $\phi_f =2 r_0^T \Omega r_\text{tot} $, which takes the analytic form
\begin{equation}
\label{eq:phase_force}
    \phi_f = \frac{4 f x_0 \delta x}{\hbar \omega} \left(e^{ t_-}  - 1 \right) \;. 
\end{equation}
The interferometer senses a force exponentially in the expansion time $t_-$. At late times, the optimal expansion time $T_o^f$ to sense $f$ is 
\begin{equation}
    T_o^f \approx - \ln \left( \frac{4 f x_0 \delta x}{\hbar \omega} \right)\;,
\end{equation}
for which $\phi_f \approx 1$. Hence, by choosing the expansion time, forces of different orders of magnitude can be sensed. 
\begin{figure*}[!t]
\includegraphics[width=0.99\textwidth]{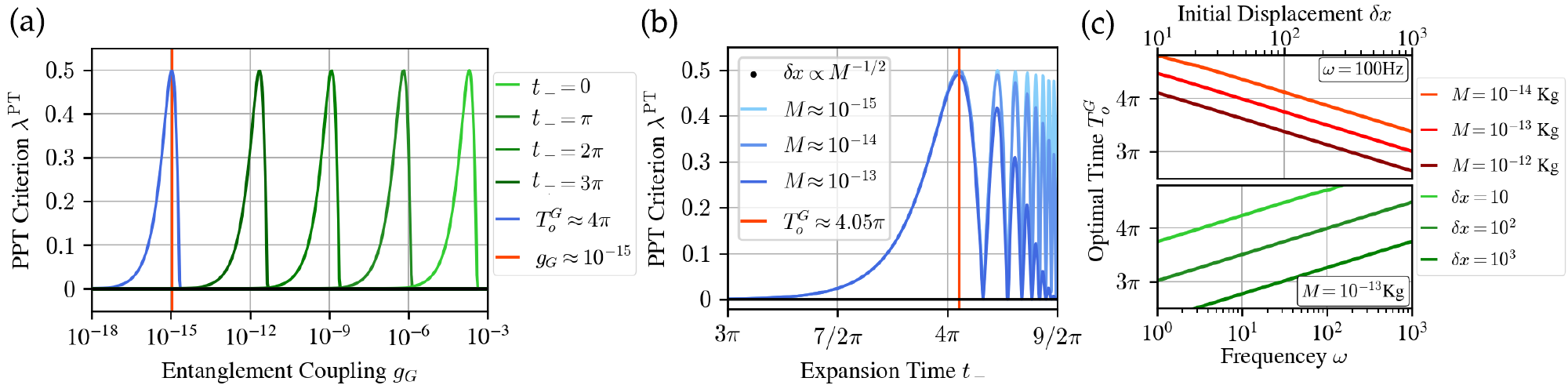}
  \caption{\label{fig:second} GIE during the expansion protocol: (a) PPT criterion for two qubits at the end of the interferometer for different expansion time as function of the coupling ($M=10^{-14}$kg, $\omega = 70$Hz, and $\delta x = 50$); (b) Time evolution of the negative eigenvalue with $\delta x \propto M^{-1/2}$; (c) the optimal time $T_o^G$ is logarithmically dependent on the initial superposition ($\delta x$) and frequency ($\omega$). }
\end{figure*}

\textit{Gravitationally Induced Entanglement} -- Two of the presented interferometers are placed at a distance $d$, and the superposition and expansion are performed collinearly with the distance $d$. If gravity is quantum in nature, the interaction is
\begin{equation}
\label{eq:hamiltoinian_grav}
    \hat{H}_G = - \frac{G M_1 M_2}{d - (\hat{X}_1 - \hat{X}_2)} \;,
\end{equation}
where $M_i$ and $X_i$ are the mass and the position operator of the $i^\text{th} = \{1,2\}$ system, measuring the displacement from the equilibrium positions of each interferometer. Such operated-valued interaction is well motivated by the quantum gravity literature~\cite{marshman_locality_2020,christodoulou_possibility_2019, bose_mechanism_2022,belenchia_quantum_2018, danielson_gravitationally_2022, carney_newton_2022,chen_quantum_2022,vinckers_quantum_2023,Chakraborty:2023kel}. We will consider $M_1 = M_2 = M$ and $d \approx 40 \mu$m -- such that the unwanted Casimir interaction can be shielded using Faraday cages~\cite{van_de_kamp_quantum_2020,schut_relaxation_2023,schut_micronsize_2023}. The interaction can be expanded up to quadratic order, and the total Hamiltonian of the two systems interacting via gravity is
\begin{equation}
\label{eq:hamiltonian_grav_expanded}
   \frac{2 \hat{H}_{\pm}^{G} }{\hbar \omega}= \sum_{i = 1,2}  \left[ \hat{p}_i^2 \pm \left( 1 \mp g_G \right) \hat{x}_i^2 \right] + 2 f_G \left(\hat{x}_1 - \hat{x}_2\right) + 2  g_G \hat{x}_1 \hat{x}_2  \;,
\end{equation}
where $\hat{x}_i$ and $\hat{p}_i$ are the unit-less position and momentum operators of the $i^\text{th}$mass, and 
\begin{equation}
\label{eq:coupling_grav}
    g_G = 2 \frac{G M }{ \omega^2 d^3} \;, \hspace{1cm} f_G =  \frac{1}{2} \frac{d}{\sqrt{2} x_0} g_G 
\end{equation}
are the dimensionless gravitational entangling coupling and gravitational force, respectively (Ref.~\cite{qvarfort2020mesoscopic,krisnanda2020observable,bose_mechanism_2022,weiss_large_2021,pedernales_enhancing_2022} and SM). In SM, the gravitational interaction for arbitrary orientations (i.e. not collinearly) is derived, showing that the chosen orientation is the optimal one (in contradiction with Ref.~\cite{nguyen_entanglement_2020,tilly_qudits_2021,schut_improving_2022, braccini_large_2023}, which assume that CS are approximately eigenstates of Eq.~\ref{eq:hamiltoinian_grav}). Furthermore, in Ref.~\cite{weiss_large_2021}, the squeezing protocol is conducted in the degree of freedom perpendicular to the displacement: also in this case, the collinear degree of freedom would be optimal.


In the SM, the dynamics is solved and the qubits joint density matrix $\rho_q$ at the end of the interferometer is derived. If gravity is not classical, at the end of the protocol, $\rho_q$ is entangled. Entanglement can be measured in terms of the Positive Partial Transpose (PPT) criterion~\cite{horodecki_quantum_2009}. For two 2-level systems, such as the studied one, a negative eigenvalue of the partially transposed density matrix ($\lambda^{\text{PT}} $) is a necessary and sufficient condition for entanglement~\cite{horodecki_separability_1996}. The witness $\mathcal{W} $ can be found by taking the partial transposition of the matrix $\ket{\lambda^{\text{PT}} } \bra{\lambda^{\text{PT}} }$, where $\ket{\lambda^{\text{PT}}}$ is the eigenvector associated to the eigenvalue $\lambda^{\text{PT}}$. Experimentally, $\mathcal{W}$ can be measured as local Pauli measurement on the two qubits (as $\mathcal{W}$ can always be decomposed in strings of Pauli matrices, which form a complete basis of the joint Hilbert space). In experimental settings, where possible measurements are restricted, other entanglement witnesses may be used~\cite{bose_spin_2017,chevalier_witnessing_2020}.

In Fig.~\ref{fig:second}, the main results of the paper are presented. By varying $t_-$, different entangling coupling strengths $g_G$  can be sensed. With $M=10^{-14}$kg, $\omega=70$Hz, and $\delta x = 70$, the interferometers sense the gravitational coupling $g_G \approx 10^{-15}$ after an expansion time $t_- \approx 4 \pi$. The presented interferometer can be used to witness entanglement induced by other forces -- such as Coulomb and Casimir which have the same expanded Hamiltonian but larger couplings~\cite{weiss_large_2021}. $\lambda^{\text{PT}}$ is found to be exponential in $t_-$ and highly oscillating after having reached its maximum. During the oscillatory phase, there is a decay in amplitude. This is due to a fundamental Humpty-Dumpty effect in sensing entanglement: the difference between the forces experienced by each branch of the wavefunction does not recombine the interferometer, leading to loss of visibility. If the initial superposition size follows $\delta x \propto M^{-1/2}$ -- which is the case of a constant force $F_q$, for instance, in diamagnetically levitated masses with dopant spins -- the time evolution of  $\lambda^{\text{PT}}$ is mass independent (for $t_- \lessapprox T_0$).
Given the three free parameters ($M$, $\omega$, $\delta x$), the optimal expansion time $T^G_o$ to sense GIE is computed -- which maximizes the PPT criterion. Similarly to the case of the force, $T^G_o$ increases logarithmically with $\omega $, and decreases logarithmically in $\delta x$ and $M$.  Among the studied parameters range, $T^G_o$ has values between $3\pi$ and $5 \pi$, placing the total protocol time in the range of $9 \pi$ and $13 \pi$. This represents an exponential improvement for GIE experiment based on cat-states sensing. 

\textit{Decoherence and Operational Precision} --  In the following, the noise analysis is presented in terms of bounds on decoherence rates and operational precision requirements, and in Appendix numerical values for different parameters ($\omega, M, \delta x$) are given. First, let us analyse the main sources of decoherence (see SM). Scattering and emission of thermal photons and variation of the center of the traps (e.g. seismic noise) are described by the Gaussian master equation
\begin{equation}
\label{eq:master_pos}
    \frac{\partial \hat{\rho}}{\partial t} = - \frac{i}{\hbar \omega} \left[ \hat{H}, \hat{\rho} \right] - \Gamma_x \left[ \hat{x}, \left[\hat{x} , \hat{\rho} \right] \right] \;,
\end{equation}
where $\Gamma_x$ is the decoherence constant in frequency units~\cite{schneider_decoherence_1999, schlosshauer_quantum_2007}. For the proposed protocol, $H = H_\pm$. The solution of this equation can be expressed in terms of the time evolution of the first and second moments, uniquely defining also the off-diagonal term of the qubit density matrix~\cite{serafini_quantum_2017, serafini_minimum_2004}. The first moment is found to be unchanged under this source of noise (and follows the symplectic transformation of Eq.~\ref{eq:symplectic_ideal}), while the second diffuses in momentum. It implies no difference in the reduced density matrix $\rho_q$ at the end of the interferometer. Hence, the expansion protocol is robust under the above source of noise. 


The Lindbladian equation describes decoherence in the qubit degrees of freedom (for $t$ in frequency units) is
\begin{equation}
\label{eq:master_spin}
    \frac{\partial \hat{\rho}_q}{\partial t} = - \frac{i}{\hbar \omega}\left[ \hat{H}_q, \hat{\rho}_q \right] - \frac{\Gamma_q}{\omega} \left[ \hat{\sigma}_z, \left[\hat{\sigma}_z , \hat{\rho}_q \right] \right] \;,
\end{equation}
where $\Gamma_q = 1/T_2$ is the qubit decoherence rate (where $T_2$ is the qubit dephasing time)~\cite{schlosshauer_quantum_2007} and $\hat{H}_q$ is the qubit free evolution Hamiltonian -- which will be ignored. Under such dissipative dynamics, the off-diagonal terms of the density matrix decay as $e^{-t \Gamma_q/\omega}$, and it can be easily generalized to the two-qubit case. Entanglement with decoherence can be measured using the same witness operator $\mathcal{W}$ from the ideal case with no decoherence. The optimal time $T_o^G$ -- which maximizes the PPT criterion -- is found to be unchanged compared to the ideal case. At this time, the entanglement is detectable for $\Gamma_q/\omega\sim \mathcal{O} (10^{-2})$. 

Additionally, given that the qubit senses a static force as in Eq.~\ref{eq:phase_force}, a random force which varies from run to run of the experiment -- usually called quasi-static noise \cite{taylor_relaxation_2007, grossardt_2020_acceleration} -- will decohere the spin. The probability distribution of such forces is assumed to be a Gaussian distribution with standard deviation $\sigma_f$. Then, the off-diagonal term of the qubit density matrix will be exponentially suppressed by a factor of 
\begin{equation}
\Gamma_f = 
     \frac{8 x_0^2 \delta x^2 \sigma_f^2}{\hbar^2 \omega^2} \left(e^{t_-} -1 \right)  \;.
\end{equation}
Then, this effect is negligible if $\Gamma_f \ll 1$, implying -- at leading order -- that $\sigma_f \ll \frac{\hbar \omega}{ 2 \sqrt{2} x_0 \delta x} e^{-t_-}$. 

Collisions with gas molecules is another source of position-localizing decoherence \cite{Romero_Isart_quantum_2011}. The decoherence rate is given by $\Gamma_\text{air} = 16 \pi \sqrt{\frac{2 \pi}{3}} \frac{P R^2}{m_\text{air} \bar{v}_\text{air}}$, where $R$ is the mass radius, $P$ the pressure, and $m_\text{air}$ and $\bar{v}_\text{air}$ are the mass and mean velocity of the molecule, respectively. It is then negligible if the total time of the protocol $t_\text{tot} \ll \omega / \Gamma_\text{mol} $, implying that
\begin{equation}
    P \ll \frac{\sqrt{3 m_\text{air} k_B T }  t_\text{tot} }{16 \pi \sqrt{\pi} \omega R^2}\;.
\end{equation}
where $k_B$ is the Boltzmann constant and $T$ is the temperature of the gas.

Experimental imperfections can arise during the protocol in the form of (a) the time taken to change from one potential to the other and (b) imprecision in the frequencies $\omega$ and switching times $t_\pm$. Regarding the former, let us define the time taken to switch potentials as $\delta t$. This effect is negligible if the sudden approximation is valid, which holds for
\begin{equation}
\label{eq:sudden_approx}
    \delta t \ll \frac{2 \sqrt{2}}{ \sqrt{2 \delta x^2+1}} e^{- 2 t_-} \; .
\end{equation}
It should be noted that this analysis also applies to Ref.~\cite{weiss_large_2021} with $\delta x=0$.

Imprecision in switching times leads to the so-called Humpty-Dumpty effect, i.e. the imperfect closure of the interferometer, inducing loss of visibility in the qubit measurements~\cite{englert_is_1988, japha_quantum_2023, xiang2024phononinducedcontrastmatter, zhou2024gyroscopicstabilitynanoparticlessterngerlach}. Errors in the frequency $\omega$ will implies the same error in the unitless switching time. This represents the dominant experimental challenge of the cat-states protocols, even without expansion. Error in the four switching times can be described as $t_i^\epsilon = t_i(1 + \epsilon_i)$, where $t_i$ are the ideal times and $\epsilon_i$ are independent and identically distributed random variables, with normal distribution $\mathcal{N}(0,\sigma_\epsilon^2)$ of standard deviation $\sigma_\epsilon \ll 1$ and zero mean (\textit{i.e.} it assumed that there are no systematic errors). The visibility as a function of $\sigma_\epsilon$ after many realizations of the experiment can be computed, by integrating the density matrix of a single run with switching error over the probability distribution of the errors. The Humpty-Dumpty effect is negligible when the visibility -- given by the norm of the off-diagonal term of $\rho_q$ -- is approximately $1$. Considering the leading term in $t_-$, this is ensured if
\begin{equation}
\label{eq:bound_humpy_dumpy}
    \sigma_\epsilon \leq \frac{4 }{\delta x \sqrt{\pi (\pi - 1)}} e^{-2 t_-} \;.
\end{equation}

The numerical values of the  bounds presented above are given in the Appendix for different masses and trap frequencies.  

\textit{Conclusions} -- Methods for creating large spatial superpositions ($\sim 100  \mu$m) of mesoscopic (micron or larger diameter) masses {\em sufficiently fast} is currently an open problem. This fastness requirement stems from necessity to minimize seismic noise, which rises at around 1 Hz, and lowering all the cooling/pressure/noise requirements for other decoherence rates, as these also need to be below the experiment's frequency. Here we remove this roadblock by showing that quantum superpositions created by weak qubit-mass forces can be enhanced exponentially fast, while interferometric completion and reading/witnessing via simple qubit measurements can be retained. While the initial small superposition is created by viable mass-qubit couplings, the mass  is decoupled from the qubit before our expansion protocol. Subsequently, the expansion is undone and the mass and qubit are coupled again to complete the interferometry. The decoupling of mass and qubit are accomplished in different ways for different qubits, but for spin qubits coupled to masses via magnetic field gradients, for example, it could be accomplished by mapping electronic spins to nuclear spins \cite{bose_spin_2017}.  A small superposition of $\delta_x = 100$ in ground-state spreads units (which can be achieved, for example, in diamagnetic traps with embedded single spin and Stern-Gerlach force for $M= 10^{-14}$kg within $0.1$s) can be exponentially expanded up to $\sim 100  \mu$m. The bounds on experimental precision and decoherence are within experimental reach. There is also a major methodological achievement: exact treatment using Gaussian dynamics is possible so that a complete wavefunction analysis of the growth of entanglement of interferometers can be made. The result not only gives force measurements with precision scaling exponentially in time -- well beyond what is called the Heisenberg limit --, but can also potentially make GIE a $\sim 10$ Hz experiment, and thereby much closer to feasibility. 

\textit{Note added:} Recent related independent work \cite{Zhao2024Integrating} has come to our attention.

\acknowledgements L.B. would like to aknowledge Engineering and Physical Sciences Research Council (EPSRC) grants (EP/R513143/1 and EP/W524335/1). S.B. would like to acknowledge EPSRC grants (EP/N031105/1, EP/S000267/1, and EP/X009467/1) and grant ST/W006227/1.

\clearpage

\appendix 

\onecolumngrid

\newpage

\begin{tabular}{ |p{6cm}|p{2cm}|p{2cm}||p{2cm}|p{2cm}|p{2cm}|p{2cm}|  }
 \hline
 \multicolumn{6}{|c|}{Comparative Analysis between Experimental Implementations} \\
 \hline
 \hline
 Parameters & Symbol & Units &  Set-up 1  & Set-up 2 & Set-up 3\\
 \hline
 \hline
 \multicolumn{6}{|c|}{Trap and Mass Parameters} \\
  \hline
 Frequency & $\omega$ & Hz   &  10   & 100 &   1000 \\
 Unit Less (UL) Initial Superposition & $\delta x$ & UL & 100  & 100 &  100 \\
 Mass & $M$ & kg &   $10^{-15}$  & $10^{-14}$ & $10^{-13}$ \\
 Density &  & kg/$\text{m}^3$ &   $3.5 \cdot 10^{3}$  &  $3.5 \cdot 10^{3}$  &  $3.5 \cdot 10^{3}$  \\
 Radius & $R$ & m &   $4.1 \cdot 10^{-7}$  & $8.8 \cdot 10^{-7}$ & $1.8 \cdot 10^{-6}$ \\
 Ground State Spread & $x_0$ & m &   $7.3 \cdot 10^{-11}$  & $7.3 \cdot 10^{-12}$ & $7.3 \cdot 10^{-13}$ \\
Initial Superposition & $\delta x x_0$ & m & $7.3 \cdot 10^{-9}$   & $7.3 \cdot 10^{-10}$  &  $7.3 \cdot 10^{-11}$  \\
 \hline
 \hline
 \multicolumn{6}{|c|}{Gravitational Induced Entanglement} \\
  \hline
 Distance between Interferometers & $d$  & m   & $40 \cdot 10^{-6}$ & $40 \cdot 10^{-6}$&  $40 \cdot 10^{-6}$\\
 Gravitational Coupling & $g_G$  & UL    &$2.1 \cdot 10^{-14}$   & $2.1 \cdot 10^{-15}$  &  $2.1 \cdot 10^{-16}$  \\
 Optimal Time & $T^{o}_G$ & s &   $1.1 $  & $0.12 $&$0.014$\\
Total Protocol Time & $t_\text{tot}$ & s& $3.2$  & $0.34$   &$0.037$  \\
Frequency of the Experiment &   & Hz &   $0.31$  & $2.9$& $27$\\
UL Maximum Superposition & $\Delta X$ &  &   $1.2 \cdot 10^{7}$  & $4.1 \cdot 10^{7}$&$1.3 \cdot 10^{8}$\\
 Maximum Superposition & $x_0 \Delta X $ & m &   $9.5\cdot 10^{-4}$  & $3.0\cdot 10^{-4}$ &$9.4\cdot 10^{-5}$ \\
\hline
 \hline
 \multicolumn{6}{|c|}{Bounds of Decoherence Rates and Operational Precision} \\
  \hline
Decoherence Rate & $\Gamma_q$ and $\Gamma_\text{air}$  & Hz &   $0.31$  & $2.8$ & $27$ \\
Gas Temperature & $T$ & K &   1  & 1 & 1 \\
Gas Pressure & $P$ & Pa &   $3.2\cdot 10^{-13}$  &  $7.5 \cdot 10^{-15}$ &  $1.7\cdot 10^{-16}$ \\
Random Force Uncertainty & $\sigma_f$ & N &   $5.6 \cdot 10^{-31}$  & $1.7 \cdot 10^{-29}$ & $5.6 \cdot 10^{-28}$ \\
 Sudden Switching Time  & $\delta t$ & s & $2.4 \cdot 10^{-13}$  & $2.4 \cdot 10^{-15}$    &$2.4 \cdot 10^{-17}$ \\
 Uncertainty in Switching Times  & $\sigma_\epsilon$ & s & $1.8 \cdot 10^{-13}$  & $1.8 \cdot 10^{-15}$   &$1.8 \cdot 10^{-17}$\\
 \hline
\end{tabular}

\section{Quantum Continuous Variables Formalism\label{app:intro}}

In this section, we review the notation and main identities used in the following and presented in the book ``\textit{Quantum Continuous Variables}"\cite{serafini_quantum_2017}.  According to the used notation, operators are labeled by $\hat{\cdot}$, while phase space quantities are not. A quantum operator is $\hat{O}$ and its expectation value is $\braket{\hat{O}} = \text{Tr} [ \hat{O} \hat{\rho} ]$, where $\hat{\rho} $ is a density matrix. The canonical operators are $\hat{x}$ and $\hat{p}$ with commutation relation $[\hat{x}, \hat{p}] = i$.  Thus, it is possible to define the following quantities
\begin{equation}
\label{eq:def_first_second}
    \hat{r} := \begin{pmatrix}
        \hat{x}\\
        \hat{p}
    \end{pmatrix} \; , \hspace{1cm}
    r := \braket{\hat{r}} = \begin{pmatrix}
        x\\
        p
    \end{pmatrix} \; , \hspace{1cm}
    \sigma := \braket{ \{ (\hat{r}-r), (\hat{r}-r)^T \} } = \begin{pmatrix}
        2\Delta_{x^2} & \Delta_{\{x,p\}} \\
        \Delta_{\{x,p\}} & 2\Delta_{p^2}
    \end{pmatrix} \;,
\end{equation}
where the first is the vector of the observables $\hat{x}$ and $\hat{p}$, the second is the vector of first moments, and the third is the second moments, also known as covariance matrix (and $\{ \cdot, \cdot \}$ represent anti-commutation), and $\Delta_A = \braket{\hat{A}^2} - \braket{\hat{A}}^2$. While Eq.~\ref{eq:def_first_second} is for a single degree of freedom, it can be easily generalized to $N$ by considering the operators $\hat{x}_i$ and $\hat{p}_i$ with $i \in \{0,..,N\}$. The most general second-order Hamiltonian can be written as
\begin{equation}
\label{eq:general_quardatic}
    \hat{H} = \frac{1}{2} \hat{r}^T H \hat{r} + \bar{r}^T \hat{r} = \frac{1}{2} \left(\hat{r} - \Tilde{r} \right)^T H \left(\hat{r} - \Tilde{r} \right) +const,
\end{equation}
where, in the case of only one degree of freedom, $H$ is a $2 \times 2$ matrix, $\bar{r}$ is a  2-dimensional vector, $\Tilde{r} = - H^{-1} \bar{r}$ (assuming an invertible $H$) and $const$ is a constant that can be disregarded. Let us now define the displacement operator and the quadratic evolution operator for time $t$ (which includes both $xp$-rotations and squeezing) as
\begin{equation}
    \hat{\mathcal{D}}_r = e^{i r^T \Omega \hat{r}} \; , \hspace{1cm} 
    \hat{S}_H = e^{- \frac{i}{2} t \hat{r}^T H \hat{r} } \;,\hspace{0.5cm} \text{where}\hspace{0.5cm} \Omega = \begin{pmatrix}
        0 & 1 \\
        -1 & 0
    \end{pmatrix} \;,
\end{equation}
is the symplectic form. Let us recall the definition of the symplectic group as the set of matrices  $S \in Sp_{2, \mathds{R}}$ such that $S \Omega S^T = \Omega$. It will also be useful to recall that the symplectic inner product is antisymmetric, i.e. $r_1^T \Omega r_2 = - r_2^T \Omega r_1 $ for any vector. By recasting the commutation relation as $[ \hat{r}, \hat{r}^T ] = i \Omega $, the action of linear and  quadratic evolutions on the canonical operators can be easily computed in the Heisenberg picture and it is given by
\begin{equation}
\label{eq:evolution_first_general}
    \hat{\cal D}^{\dag}_r \hat{r} \hat{\cal D}_r = \hat{r}-r,
    \hspace{0.5cm}
    \hat{S}_H \hat{r} \hat{S}_H^{\dag} = S_H^{-1} \hat{r},\hspace{0.5cm} \text{where}\hspace{0.5cm}  S_H = e^{t \Omega H}\;,
\end{equation}
it the symplectic transformation associated with $H$
All of these formulae could be recast in terms of ladder operators $\hat{a} = \frac{\hat{x} + i \hat{p}}{\sqrt{2}}$, which lead to $\hat{\mathcal{D}}_r = \hat{\mathcal{D}}_{-\alpha} = e^{\alpha^* \hat{a} - \alpha \hat{a}}$, where $\alpha = \frac{x + i p}{\sqrt{2}} $. A QHO and an IHO give the symplectic transformations
\begin{equation}
\label{eq:simpletic_transf_important}
    S_{H_+} = e^{- t_+ \Omega H_+ } = \begin{pmatrix}
        \cos t_+ & - \sin t_+ \\
        \sin t_+ & \cos t_+
    \end{pmatrix} \;, \;\;\; \text{and} \;\;\; S_{H_-}= e^{- t_- \Omega H_- } = \begin{pmatrix}
        \cosh t_- & - \sinh t_- \\
        - \sinh t_- & \cosh t_-
    \end{pmatrix} \;,
\end{equation}
respectevely. The following identities will be useful
\begin{equation}
\label{eq:addition_disp}
\hat{\mathcal{D}}_{r_1}^\dag = \hat{\mathcal{D}}_{-r_1}, \hspace{1cm}
    \hat{\mathcal{D}}_{r_1} \hat{\mathcal{D}}_{r_2} = \hat{\mathcal{D}}_{r_1+r_2} e^{-\frac{i}{2} r_1^T \Omega r_2} \hspace{0.5cm} \implies \hspace{0.5cm}   \hat{\mathcal{D}}_{-r_1} \hat{\mathcal{D}}_{r_2} \hat{\mathcal{D}}_{ r_1} = \hat{\mathcal{D}}_{r_2} e^{ i r_1^T \Omega r_2} \;,
\end{equation}
\begin{equation}
\label{eq:sim_disp_com}
    \hat{S}_H \hat{\mathcal{D}}_{r} = \hat{\mathcal{D}}_{S_H^{-1} r} \hat{S}_H \hspace{0.5cm} \implies \hspace{0.5cm}   \hat{S}_H^\dag \hat{\mathcal{D}}_{r} = \hat{\mathcal{D}}_{S_H r} \hat{S}_H^\dag \;.
\end{equation}

Then, the unitary generated by the Hamiltonian of Eq.~\ref{eq:general_quardatic} can be rewritten as
\begin{equation}
    \hat{U}(t) = e^{i \hat{H} t} = \hat{\mathcal{D}}_{\tilde{r}}^\dag e^{\frac{i}{2} t \hat{r}^T H \hat{r}   } \hat{\mathcal{D}}_{ \tilde{r}} = \hat{\mathcal{D}}_{-\tilde{r}} \hat{S}_H(t) \hat{\mathcal{D}}_{ \tilde{r}} = {\rm e}^{\frac{i}{2}\tilde{r}S_H^{T}\Omega\tilde{r}}\hat{\mathcal{D}}_{\left(  S_H^{-1} - \mathds{1} \right)\tilde{r}} \hat{S}_H (t)\; .
\end{equation}

\section{Creation of Cat State and Expansion Protocol}
Let us apply this formalism to prove Eq.~\ref{eq:cat_state}. The total Hamiltonian of a trapped mass of frequency $\omega$ and ground spread $x_0$ and a qubit of frequency $\omega_0$, interacting with an operated value force $F_q$ is given by:
\begin{equation}
    \frac{\hat{H}_{\text{mq}}}{\hbar} = \omega_0 \hat{\sigma}_z + \frac{\omega}{2} \left(\hat{x}^2 + \hat{p}^2 \right) - F_q x_0 \hat{x} \otimes \sigma_z  =  \omega_0 \hat{\sigma}_z + \frac{\omega}{2} \left(\left( \hat{x} - \frac{\delta x}{2} \hat{\sigma}_z\right)^2 + \hat{p}^2 \right) + const.
\end{equation}
where the constant is from $\sigma_z^2 = \mathds{1}$ and it can be ignored. Let us define the vector $\tilde{r} =\delta x/2 \left( 1 \;\;\; 0 \right)^T$. The unitary -- up to an overall phase -- can be rewritten in the form
\begin{equation}
    \hat{U}_{\text{mq}} = e^{-i \omega_0 \sigma_z t_0} \hat{\mathcal{D}}_{- \tilde{r} \sigma_z } \hat{S}_{H_+} \hat{\mathcal{D}}_{ \tilde{r} \sigma_z } = e^{i \omega_0 \sigma_z t} \hat{\mathcal{D}}_{ \left( S^{-1}_{H_+}  - \mathds{1} \right) \tilde{r} \sigma_z }  \hat{S}_{H_+} \;, \;\;\; \text{where in this case} \;\;\; S_{H_+} = e^{- t_0 \Omega H_+ } \;.
\end{equation}
By noticing that $ \hat{S}_{H_+} \ket{0}_m = \ket{0}_m$, the state $\frac{1}{\sqrt{2}}\ket{0}_m \otimes (\ket{+1}_q+\ket{-1}_q)$ evolves at time $t_0$ to
\begin{equation}
    \ket{\psi(t_0)} = \frac{1}{\sqrt{2}}\left(e^{-i\phi(t_0)} \ket{r_0 (t_0)}_m \ket{+1}_q + e^{+i\phi(t_0)} \ket{-r_0(t_0)}_m \ket{-1}_q \right)\;, \;\;\; \text{where} \;\;\; r_0(t_0) = \frac{\delta x}{2} \begin{pmatrix}
        1- \cos t_0 \\
        - \sin t_0
    \end{pmatrix} \;,
\end{equation}
where $\ket{r}:= \hat{\mathcal{D}}_r \ket{0}$ is a Coherent State (CS) and $\phi(t_0) = \omega_0 t_0$ is the trivial dynamical phase of the qubit and will be omitted in the following.  At $t_0=\pi$, Eq.~\ref{eq:cat_state} is recovered. 

The expansion and recombination protocol is given by the unitary in Eq.~\ref{eq:unitary_ideal}. As claimed, the total phase-space transformation is given by a symplectic transformation, as Eq.~\ref{eq:unitary_ideal} is only quadratic. It is given by, 
\begin{equation}
    S_{\text{tot}}^{-1} = S_{H_-}^{-1} S_{H_+}^{-1} S_{H_-}^{-1} S_{H_+}^{-1} = S_{H_-}^{-1} \Omega \left(S_{H_-}^{-1}\right)^T \Omega = - \mathds{1} \;,
\end{equation}
where $S_{H_\pm}$ are given in Eq.~\ref{eq:simpletic_transf_important}. Similarly, it is possible to construct the symplectic transformation $S(t)$ at time $t$ of the protocol. For instance, the maximum expansion is achieved at $t_\text{max} = t_- + 3 \pi/4 $ and the transformation is given by
\begin{equation}
   S(t_\text{max})^{-1} =  \left(S_{H_+}\left(t_+ = \pi/4 \right) \right)^{-1}  S_{H_-}^{-1} S_{H_+}^{-1} = \frac{1}{\sqrt{2}} \begin{pmatrix}
       1 & 1 \\ -1 & 1
   \end{pmatrix} S_{H_-}^{-1} S_{H_+}^{-1} = 
   \frac{1}{\sqrt{2}} \begin{pmatrix}
       -e^{t_-} & e^{t_-} \\ - e^{-t_-} & - e^{-t_-}
   \end{pmatrix}
\end{equation}
Hence, 
\begin{equation}
    r_\pm (t_\text{max}) =  S(t_\text{max})^{-1} \begin{pmatrix}
        \pm \delta x \\
        0
    \end{pmatrix} =
    \mp \frac{\delta x}{\sqrt{2}}
    \begin{pmatrix}
        e^{t_-} \\
        e^{-t_-}
    \end{pmatrix}
\end{equation} 
Then, the maximum superposition $\Delta x = \left|r_+(t_\text{max}) - r_-  (t_\text{max})  \right|_1 =  \sqrt{2} e^{t-} \delta x $, as claimed, while the difference in momentum exponentially decays in $t_-$.

\section{Dynamics with External Force and Entanglement: Derivation of the Form $\hat{U} = \hat{\mathcal{D}}_{r_{\text{tot}}} \hat{S}_\text{tot} $ \label{app:unitaries}}

The Hamiltonians of the protocol with the addition of a force read
\begin{equation}
    \hat{H}_\pm^f = \frac{\hat{P}^2}{2 M} \pm \frac{1}{2} M \omega^2 \hat{X}^2 - f \hat{X} = \frac{\hbar \omega}{2} \left( \hat{p}^2 \pm \hat{x}^2 - 2 g \hat{x} \right) = \frac{\hbar \omega}{2} \left( \hat{p}^2 \pm (\hat{x}_g^\mp)^2 \right)+const ,
\end{equation} 
where $g = f x_0/(\hbar \omega)$ and we are defining the shifted operators $x_g^{\pm} = x \pm g$, which preserve the algebra. Hence, it follows that $ \hat{S}_{H_\pm}^f = \exp\left( i \hat{H}_\pm^f t_\pm \right) = \hat{\mathcal{D}}_{\mp r_g} \hat{S}_{H_\pm} \hat{\mathcal{D}}_{\pm r_g} $, where $r_g = (g \;\;0)^T$. The total unitary of the protocol with force becomes
\begin{equation}
    U^f =  \hat{S}_{H_-}^f\hat{S}_{H_+}^f\hat{S}_{H_-}^f \hat{S}_{H_+}^f = \hat{\mathcal{D}}_{ r_g} 
      \hat{S}_{H_-} \hat{\mathcal{D}}_{- 2 r_g}
      \hat{S}_{H_+} \hat{\mathcal{D}}_{2 r_g} 
      \hat{S}_{H_-}\hat{\mathcal{D}}_{- 2 r_g} \hat{S}_{H_+} \hat{\mathcal{D}}_{ r_g} \;.
\end{equation}
By repetitive use of Eq.~\ref{eq:sim_disp_com}, it follows that
\begin{equation}
    U^f =  \hat{S}_{H_-}^f\hat{S}_{H_+}^f\hat{S}_{H_-}^f \hat{S}_{H_+}^f = \hat{\mathcal{D}}_{ r_g} 
       \hat{\mathcal{D}}_{- 2 S^{-1}_{H_-} r_g}
      \hat{\mathcal{D}}_{2 S^{-1}_{H_-} S^{-1}_{H_+} r_g} 
      \hat{\mathcal{D}}_{- 2 S^{-1}_{H_-} S^{-1}_{H_+} S^{-1}_{H_-} r_g} \hat{\mathcal{D}}_{S^{-1}_{H_-} S^{-1}_{H_+} S^{-1}_{H_-} S^{-1}_{H_+}  r_g} \hat{S}_{H_-} \hat{S}_{H_+} \hat{S}_{H_-} \hat{S}_{H_+}.
\end{equation}
The total unitary can be rewritten in the form $U^f = {\rm e}^{i\varphi}\hat{\mathcal{D}}_{r_\text{tot}} U_{\text{tot}}$, where ${\rm e}^{i\varphi}$ is an irrelevant overall phase (which will not be computed), and
\begin{equation}
    r_\text{tot} = r_g - 2  S^{-1}_{H_-} r_g + 2  S^{-1}_{H_-}  S^{-1}_{H_+} r_g - 2 S^{-1}_{H_-}  S^{-1}_{H_+} S^{-1}_{H_-}  r_g + S^{-1}_{H_-}  S^{-1}_{H_+} S^{-1}_{H_-} S^{-1}_{H_+}  r_g \;.
\end{equation}
By using  $S_{H_+}^{-1} = \Omega  $ at $t_+=\frac\pi2$ and $S_{H_-}=S_{H_-}^{T} \; \forall t_-$, one has  $S^{-1}_{H_+}  S^{-1}_{H_-} S^{-1}_{H_+} S^{-1}_{H_-} = -\mathds{1}$ and $\Omega^{-1} = - \Omega$. Furthermore, from the definition of sympaletic group $S^{-1}_{H_-}  S^{-1}_{H_+} S^{-1}_{H_-} = - \Omega $,  such that \begin{equation}
\label{eq:phase_symplectic}
    r_\text{tot} = - 2 S_{H_-}^{-1} r_g + 2 S_{H_-}^{-1}   \Omega r_g + 2  \Omega  r_g = 
\begin{pmatrix}
 - 2 g e^{t_-} \\
 -2 g \left(e^{t_-}+1\right) \\
\end{pmatrix}
\;.
\end{equation}

Let us now move to the case of the entangling dynamics via gravity. The expanded Hamiltonian is given in Eq.~\ref{eq:hamiltonian_grav_expanded}, which contains both linear and quadratic terms.  It can be rewritten in the form $\hat{H}_\pm^{G} = \frac{1}{2}\left(\hat{r} - r_\pm\right)^T H_\pm^G \left(\hat{r} - r_\pm\right) $ with 
\begin{equation}
    \frac{ H_\pm^G}{\hbar \omega} = \begin{pmatrix}
        \pm \left(1 \mp g_G \right) & 0 & g_G & 0 \\
         0 & 1 & 0 & 0 \\
         g_G  & 0 & \pm \left(1 \mp g_G \right) & 0 \\
         0 & 0 & 0 & 1
    \end{pmatrix} \; , \hspace{1cm}
    r_\pm = - \left( H_\pm^G \right)^{-1} \begin{pmatrix}
        f_G \\
          0 \\
         -f_G \\
         0
    \end{pmatrix}   = \pm \frac{ g_G f_G}{2g_G \mp 1} \begin{pmatrix}
        1 \\
          0 \\
         -1 \\
         0
    \end{pmatrix}   \;,
\end{equation}
where $f_G$ and $g_G$ are given in Eq.~\ref{eq:coupling_grav}. Then, $ e^{i \hat{H}_\pm^{G} t_\pm } = \hat{\mathcal{D}}_{- r_\pm} \hat{S}_{H_\pm^G}  \hat{\mathcal{D}}_{ r_\pm} $, where we defined the entangling quadratic term $\hat{S}_{H_\pm^{G}} = \exp \left( \frac{i}{2} \hat{r}^T H_\pm^G  \hat{r}\right) $. Then the total unitary can be rewritten as:
\begin{align}
    U^G &= e^{i \hat{H}_-^{G} t_- } e^{i \hat{H}_+^{G} t_+ } e^{i \hat{H}_-^{G} t_- } e^{i \hat{H}_+^{G} t_+ } \\ 
    &= \hat{\mathcal{D}}_{- r_-} \hat{S}_{H_-^G}  \hat{\mathcal{D}}_{ \left( r_- - r_+\right)} \hat{S}_{H_+^G}  \hat{\mathcal{D}}_{ \left( r_+- r_-\right)} \hat{S}_{H_-^G}  \hat{\mathcal{D}}_{ \left( r_-- r_+\right)} \hat{S}_{H_+^G}  \hat{\mathcal{D}}_{  r_+} \;.
\end{align}
Applying again the same procedure done in the case of the force, the final displacement is given by
\begin{equation}
    r_{\text{tot}} = - r_- + S^{-1}_{H_-^G} \left( r_- - r_+\right) - S^{-1}_{H_-^G} S^{-1}_{H_+^G} \left( r_- - r_+\right) + S^{-1}_{H_-^G} S^{-1}_{H_+^G}  S^{-1}_{H_-^G} \left( r_- - r_+\right) + S^{-1}_{H_-^G} S^{-1}_{H_+^G} S^{-1}_{H_-^G} S^{-1}_{H_+^G} r_+ \;,
\end{equation}
where we inroduced the symplectic transformation $S_{H_\pm^G} = \exp \left( \Omega H_\pm^G t_\pm \right)$. Thus, the unitary is $U^G = \hat{\mathcal{D}}_{r_\text{tot}} \hat{S}_{\text{tot}}^G$ where $\hat{S}_{\text{tot}}^G = \hat{S}_{H_-^G}\hat{S}_{H_+^G}\hat{S}_{H_-^G}\hat{S}_{H_+^G}$.

\section{Derivation of Reduced Density Matrices \label{app:reduced}}

In this section, the reduced density matrix of the qubit(s) at the end of the interferometer is derived. We present both the case of a single interferometer and the case of two interferometers undergoing entangling dynamics.

The positron degree of freedom firstly experiences a qubit-dependent displacement ($\hat{\mathcal{D}}_{\pm r_{0}} $), then undergoes the Gaussian evolution ($U = \hat{\mathcal{D}}_{r_{\text{tot}}} S_\text{tot} $) and, finally, is displaced back ($\hat{\mathcal{D}}_{\pm r_{0}} $). The total density matrix of the mass-qubit system at the end of the interferometer is given by
\begin{equation}
\hat{\rho}(t_f) = \frac{1}{2} \sum_{j,k = \pm 1} \hat{\rho}_{j,k} \otimes \ket{j} \bra{k}\;, \;\;\; \text{where} \;\;\;
    \hat{\rho}_{j,k} = \hat{\mathcal{D}}_{j r_0}  \hat{\mathcal{D}}_{r_{\text{tot}}} \hat{S}_\text{tot} \hat{\mathcal{D}}_{j r_0} \ket{0}\bra{0} \hat{\mathcal{D}}_{- k r_0}  \hat{S}_\text{tot}^\dag \hat{\mathcal{D}}_{- r_{\text{tot}}}  \hat{\mathcal{D}}_{- k r_0}  \;,
\end{equation}
where we used the fact that $ \hat{\mathcal{D}}_{r}^\dag =\hat{\mathcal{D}}_{-r} $. The reduced density matrix can be computed by tracing out the continuous variable, which gives the coefficients
\begin{align}
    \rho_{j,k} &= \text{Tr} \left[ \ket{0}\bra{0} \hat{\mathcal{D}}_{- k r_0}  \hat{S}_\text{tot}^\dag \hat{\mathcal{D}}_{- r_{\text{tot}}}  \hat{\mathcal{D}}_{- k r_0}\hat{\mathcal{D}}_{j r_0}  \hat{\mathcal{D}}_{r_{\text{tot}}} \hat{S}_\text{tot} \hat{\mathcal{D}}_{j r_0}   \right] \\ 
     &=\text{Tr} \left[ \ket{0}\bra{0} \hat{S}_\text{tot}^\dag  \hat{\mathcal{D}}_{- k S^{-1}_\text{tot}  r_0}  \hat{\mathcal{D}}_{ - r_{\text{tot}}}  \hat{\mathcal{D}}_{- k r_0} \hat{\mathcal{D}}_{j r_0}  \hat{\mathcal{D}}_{r_{\text{tot}}} \hat{\mathcal{D}}_{ j S^{-1}_\text{tot}   r_0}  \hat{S}_\text{tot}  \right] \\
     &= \text{Tr} \left[ \ket{0}\bra{0} \hat{S}_\text{tot}^\dag  \hat{\mathcal{D}}_{ - r_{\text{tot}} - k \left(\mathds{1} + S^{-1}_\text{tot}  \right) r_0 }  \hat{\mathcal{D}}_{ r_{\text{tot}} + j \left(\mathds{1} + S^{-1}_\text{tot}  \right) r_0} \hat{S}_\text{tot}  \right] e^{ \frac{i}{2} (j - k) r_\text{tot}^T\Omega \left( \mathds{1} - S^{-1}_\text{tot}\right) r_0 } \\
     &= \text{Tr} \left[ \ket{0}\bra{0} \hat{S}_\text{tot}^\dag  \hat{\mathcal{D}}_{ \left( j - k \right) \left(\mathds{1} + S^{-1}_\text{tot}  \right) r_0 } \hat{S}_\text{tot}  \right] e^{\frac{i}{2} (j - k) r_\text{tot}^T\Omega \left[  \left( \mathds{1} - S^{-1}_\text{tot}\right) + \left( \mathds{1} + S^{-1}_\text{tot}\right) \right]r_0 } \\
     &= \text{Tr} \left[ \ket{0}\bra{0}  \hat{\mathcal{D}}_{ \left( j - k \right) S_\text{tot} \left(\mathds{1} + S^{-1}_\text{tot}  \right) r_0 } \right] e^{ i (j - k) r_\text{tot}^T\Omega r_0 }
\end{align}
where the first equality follows from the cyclicity of the trace; the second from Eq.~\ref{eq:sim_disp_com} and its conjugate; the third and forth from repetitively using Eq.~\ref{eq:addition_disp} and noticing that $k^2 (S^{-1}_\text{tot}  r_0)^T \Omega r_0 = - j^2  r_0^T \Omega S^{-1}_\text{tot} r_0  $ for $j,k = \pm1$. Let us recall that $\braket{0|r} = \exp(- \frac{1}{4} ||r||^2)$  -- where $|| \cdot || $ stands for the vector norm. Then, noticing that
\begin{equation}
    || S_{\text{tot}} r ||^2  = r^T  S_{\text{tot}}^T \Omega^T \Omega S_{\text{tot}} r = r^T \Omega^T S^{-1}_{\text{tot}}     S^{-1 T}_{\text{tot}} \Omega r   = r^T \Omega^T \sigma(t_f) \Omega r ,
\end{equation}
we get to the one of our main results:
\begin{equation}
\label{eq:density_general_1qubit}
   \rho_{j,k} = \exp \left[ - \frac{1}{2} (1 - j k) r_0^T \left(\mathds{1} + S^{-1}_\text{tot}  \right)^T \Omega^T \sigma(t_f) \Omega \left(\mathds{1} + S^{-1}_\text{tot}  \right) r_0 + i (j - k) r_\text{tot}^T\Omega r_0   \right] ,
\end{equation}
Recalling that $r(t) =  S^{-1}_\text{tot} r_0$, the matrix form of the qubit at time $t_f$ is then given
\begin{equation}
\label{eq:density_general_1qubit_matrix}
    \hat{\rho}_R (t_f)= \frac{1}{2} \begin{pmatrix}
        1 & e^{ -   \left(r_0 + r(t_f) \right)^T \Omega^T \sigma(t_f) \Omega \left(r_0 + r(t_f) \right) - 2 i r_0^T \Omega r_{\text{tot}} }\\
        e^{-   \left(r_0 + r(t_f) \right)^T \Omega^T \sigma(t_f) \Omega \left(r_0 + r(t_f) \right) + 2 i r_0^T \Omega r_{\text{tot}} } & 1
    \end{pmatrix}
\end{equation}
In the case of $S_\text{tot} \neq - \mathds{1} $, the first term in the exponentials is the Humpty-Dumpty effect, which makes the contrast exponentially decay. For $S_\text{tot} = - \mathds{1}$, then $r(t_f) = -r_0$ and $\sigma(t_f) = \mathds{1} $, such that
\begin{equation}
    \hat{\rho}_\text{tot} (t_f)= \frac{1}{2} \begin{pmatrix}
        1 & e^{ - 2 i r_0^T \Omega r_{\text{tot}} }\\
        e^{ 2 i r_0^T \Omega r_{\text{tot}} } & 1
    \end{pmatrix}
\end{equation}
which is a pure state, so that the final spin state is given by $\ket{\psi}_S = \frac{1}{2} (\ket{0} + e^{ i \phi_f} \ket{1} )$, as claimed. This can be easily computed in the specific case of $S_\text{tot} = - \mathds{1}$:
\begin{align}
    \rho_{j,k} &= \text{Tr} \left[ \ket{0}\bra{0} \hat{\mathcal{D}}_{- k r_0}  \hat{S}_\text{tot}^\dag \hat{\mathcal{D}}_{- r_{\text{tot}}}  \hat{\mathcal{D}}_{- k r_0}\hat{\mathcal{D}}_{j r_0}  \hat{\mathcal{D}}_{r_{\text{tot}}} \hat{S}_\text{tot} \hat{\mathcal{D}}_{j r_0}   \right] \\ 
     &= \text{Tr} \left[ \ket{0}\bra{0} \hat{S}_\text{tot}^\dag  \hat{\mathcal{D}}_{k r_0}  \hat{\mathcal{D}}_{- r_{\text{tot}}}  \hat{\mathcal{D}}_{- k r_0} \hat{\mathcal{D}}_{j r_0}  \hat{\mathcal{D}}_{r_{\text{tot}}} \hat{\mathcal{D}}_{- j  r_0}  \hat{S}_\text{tot}  \right] \\
     &= e^{i \left( k - j \right) r_0^T \Omega r_{\text{tot}}} \text{Tr} \left[ \ket{0}\bra{0} \hat{S}_\text{tot}^\dag   \hat{\mathcal{D}}_{- r_{\text{tot}}}   \hat{\mathcal{D}}_{r_{\text{tot}}} \hat{S}_\text{tot}  \right] = e^{i \left( k - j \right) r_0^T \Omega r_{\text{tot}}}
\end{align}

Moving on to entanglement, the joint density matrix of the two mass-qubit systems at the end of the interferometers is given by
\begin{equation}
\hat{\rho}(t_f) = \frac{1}{4} \sum_{j,k,m,n \in \{-1,1\}} \hat{\rho}_{j,k}^{m,n} \otimes \ket{j} \bra{m} \otimes \ket{k} \bra{n} 
\end{equation}
where the 16 density matrices of the two degrees of freedom are
\begin{equation}
     \hat{\rho}_{j,k}^{m,n}  = \hat{\mathcal{D}}_{r_{j,k}}  \hat{\mathcal{D}}_{r_{\text{tot}}} \hat{S}^G_\text{tot} \hat{\mathcal{D}}_{r_{j,k}} \ket{0,0}\bra{0,0} \hat{\mathcal{D}}_{- r_{m,n}}  \left( \hat{S}^G_\text{tot}\right)^\dag \hat{\mathcal{D}}_{- r_{\text{tot}}}  \hat{\mathcal{D}}_{- r_{m,n}} 
\end{equation}
where we defined the joint ground state of the two systems $\ket{0,0}$ and the four-dimensional vectors 
\begin{equation}
    r_{j,k} = \left( j x_0 \;\;\; j p_0 \;\;\; k x_0 \;\;\; k p_0 \right)^T
\end{equation} 
with $j,k = \pm1$. By tracing out the positions Hilbert spaces, the coefficients of the spins joint dentistry matrix are
\begin{align}
    \rho_{j,k}^{m,n} &= \text{Tr} \left[ \ket{0,0}\bra{0,0}   \hat{\mathcal{D}}_{- r_{m,n}}  \left( \hat{S}^g_\text{tot}\right)^\dag \hat{\mathcal{D}}_{- r_{\text{tot}}}  \hat{\mathcal{D}}_{- r_{m,n}}   \hat{\mathcal{D}}_{r_{j,k}}  \hat{\mathcal{D}}_{r_{\text{tot}}} \hat{S}^g_\text{tot} \hat{\mathcal{D}}_{r_{j,k}} \right] \\ 
    &= \text{Tr} \left[ \ket{0,0}\bra{0,0} \left( \hat{S}^g_\text{tot}\right)^\dag   \hat{\mathcal{D}}_{- (S^g_\text{tot})^{-1} r_{m,n}}  \hat{\mathcal{D}}_{- r_{\text{tot}}}  \hat{\mathcal{D}}_{- r_{m,n}}   \hat{\mathcal{D}}_{r_{j,k}}  \hat{\mathcal{D}}_{r_{\text{tot}}} \hat{\mathcal{D}}_{(S^G_\text{tot})^{-1} r_{j,k}} \hat{S}^G_\text{tot} \right] \\ 
    &= e^{- \frac{i}{2} \Phi_{j,k}^{m,n} } \text{Tr} \left[ \ket{0,0}\bra{0,0} \left( \hat{S}^G_\text{tot}\right)^\dag   \hat{\mathcal{D}}_{\left(\mathds{1} +  (S^G_\text{tot})^{-1} \right) \left(r_{j,k} -  r_{m,n} \right)}  \hat{S}^G_\text{tot} \right] \\ 
    &= e^{- \frac{i}{2} \Phi_{j,k}^{m,n} } \bra{0,0} \hat{\mathcal{D}}_{\left(\mathds{1} +  S^G_\text{tot} \right) \left(r_{j,k} -  r_{m,n} \right)}  \ket{0,0} 
\end{align}
where we recall that in this case $S^G_\text{tot} \neq -\mathds{1} $. The phases can be computed by the repeat use of summation of displacement operators and they are given by 
\begin{equation}
    \Phi_{j,k}^{m,n} = \left[\left(S_{ \text{tot}}^{G}\right)^{-1} \left( r_{j,k} - r_{m,n} \right) \right]^T   \Omega \left(r_{m,n} + r_{j,k}\right) +  2 \left( r_{j,k} - r_{m,n} \right)^T \Omega r_{\text{tot}} 
\end{equation}
The joint density matrix elements are
\begin{equation}
\label{eq:density_general_2}
     \rho_{j,k}^{m,n} = \exp{\left[- \frac{1}{4} || \left( S_{\text{tot}}^{G}  + \mathds{1}  \right)\left(r_{m,n} - r_{j,k} \right)||^2  - \frac{i}{2} \Phi_{j,k}^{m,n}   \right]} \; . 
\end{equation} 
The first terms of the exponent is the fundamental Humpty-Dumpty effect due to different accelerations for each arm of the interferometer. The second terms are the phases for detecting entanglement. 

\section{Expansion of Gravitational Interaction to Quadratic Form}

In the low-energy regime, the gravitational interaction between two masses is given by the Newtonian potential. The centers of the two interferometers are taken at $(0,0,0)^T$ and $(d \cos\theta, d \sin \theta, 0)^T$.  The position operators $\hat{X}_i$ of each mass measure the displacement between from the center in the $x$-direction (and the displacements in the other directions will be omitted). Then, the operator-valued distance between the masses is given by 
\begin{equation}
    \hat{\mathbf{d}} =( (\hat{X}_1 - \hat{X}_2) -  d \cos\theta, \;\; - d \sin \theta, \;\;  0)^T
\end{equation}
By computing the norm $\left| \hat{\mathbf{d}} \right|$, the Hamiltonian reads
\begin{align}
    \hat{H}_G \left( \theta \right) &=-  \frac{G M_1 M_2}{d \sqrt{1 + \left(\frac{\hat{X}_1 - \hat{X}_2}{d} \right)^2 - 2 \cos \theta \left( \frac{\hat{X}_1 - \hat{X}_2}{d} \right)}} \\ 
    &= - \frac{G M_1 M_2}{d} \left(1 + \frac{\cos \theta}{d} \left( \hat{X}_1 - \hat{X}_2 \right) + \frac{1 + 3 \cos 2 \theta}{4 d^2} \left( \hat{X}_1 - \hat{X}_2 \right)^2  \right) + \mathcal{O} \left( \left( \frac{\hat{X}_1 - \hat{X}_2}{d} \right)^3 \right)
\end{align}
where in the last inequality the potential was Taylor expanded for small displacements from the center, \textit{i.e.} $d \gg \braket{\hat{X}_1 - \hat{X}_2}$. By moving to the unit-less operator $\hat{X}_i = \sqrt{2} x_0 \hat{x}_i$, considering $M_1= M_2 = M$, and dividing by $\hbar \omega / 2$, the interaction is
\begin{align}
    \hat{H}_G \left( \theta \right) = - 2 f_G(\theta) \left( \hat{x}_1 - \hat{x}_2 \right) - g_G(\theta) \left( \hat{x}_1 - \hat{x}_2 \right)^2   + const.
\end{align}
which recovers Eq.~\ref{eq:hamiltonian_grav_expanded}, and where,
\begin{equation}
    f_G(\theta) = \frac{G M}{\omega^2 d^3}  \frac{d}{\sqrt{2}x_0} \cos \theta \hspace{1cm}
     g_G(\theta) =  \frac{G M}{2 \omega^2 d^3} \left(1 + 3 \cos 2 \theta \right)
\end{equation}
The linear set up $\theta = 0$ gives
\begin{equation}
    f_G = \frac{G M}{\omega^2 d^3}  \frac{d}{\sqrt{2}x_0} \hspace{1cm }  g_G = 2 \frac{G M}{\omega^2 d^3} 
\end{equation}
which implies Eq.~\ref{eq:coupling_grav}. For the parallel case ($\theta = \pi/2$), $f_G \to 0$ and $g_G  \to -\frac{1}{2} g_G $.  This recovers the result of Ion trap literature, where the coulomb force can be used as beam splitter interaction \cite{rojokov_two_2024}. 

\section{Phase Space Methods: Characteristic and Wigner Functions}

The set of displacement operators $\hat{\mathcal{D}}_r$ forms a complete basis such that any density matrix can be written as
\begin{equation}
    \rho = \frac{1}{(2 \pi )^n}\int_{\mathbb{R}^{2 n}} d r \chi (r) \hat{\mathcal{D}}_r \; , \;\;\; \text{where} \;\;\;
     \chi (r) = \text{Tr} \left[ \rho \hat{\mathcal{D}}_{-r} \right]
\end{equation}
is the characteristic function, and $n$ is the number of degrees of freedom. The characteristic and Wigner functions of any time-dependent Gaussian state -- with first and second moments $r(t)$ and $\sigma (t)$ -- are given by
\begin{equation}
    \chi(\bar{r},t) = e^{-\frac{1}{4} \bar{r}^T \Omega^T \sigma (t)  \Omega \bar{r} + i \bar{r}^T \Omega^T r(t)} \;, \hspace{1cm}
    \mathcal{W} (\bar{r},t) = \frac{2^n}{\pi^n \sqrt{\text{det} \sigma(t) }}e^{-\left(\bar{r} - r(t) \right)^T  \sigma (t) ^{-1} \left(\bar{r} - r(t) \right)} \;,
\end{equation}
respectively \cite{serafini_quantum_2017}.  Hence, under a Gaussian evolution $\hat{U}_{\text{tot}}(t)$ only quadratic in $\hat{x}$ and $\hat{p}$  (Eq.~\ref{eq:unitary_ideal}) with associated symplectic transformation $S(t)$, the initial coherent states with $r_\pm(0) = \pm r_0$ and $\sigma_\pm(0) = \mathds{1} $ have the characteristic functions is
\begin{equation}
\label{eq:characterstic_time}
    \chi_{\pm}(\bar{r},t) = e^{- \frac{1}{4} \bar{r}^T \Omega^T S^{-1}_\text{tot}(t)  \left(S^{-1}_\text{tot}(t) \right)^T \Omega \bar{r} \pm i \bar{r}^T \Omega^T  S^{-1}_\text{tot}(t)r_0 }
\end{equation}
The characteristic function of the whole quantum state during the protocol can be found by tracing out also the qubit Hilbert space, such that $\chi_{\text{tot}}(r,t) = \frac{1}{2} \left( \chi_{+}(r,t) + \chi_{-}(r,t) \right)$. The same will apply to the case of Wigner functions (Fig.~\ref{fig:first}). It is then easy to check that $\text{Tr}( \rho(t)) = \chi(0,t) = 1$. Expectation values can be computed as $\braket{\hat{r}^n}= \left. \left(-i \frac{\partial}{\partial (\Omega r)} \right)^n \chi(r) \right|_{r=0}$.  For the analysis in Sec.~\ref{sm:linearity_switching}, the following identities are needed 
\begin{equation}
\label{eq:exp_x3x4}
    \braket{\hat{x}^3(t)} = \left. - i   \frac{\partial^3 }{\partial p^3}  \chi (\bar{r}, t) \right|_{\bar{r}=0}  =0 \hspace{1cm}  \braket{\hat{x}^4(t)} =  \left. \frac{\partial^4 }{\partial p^4} \chi (\bar{r}, t) \right|_{\bar{r}=0} = \frac{3}{4} \left(\Delta_{x^2}\right)^2 + 3  \Delta_{x^2}  x^2 + x^4
\end{equation}
\begin{equation}
\label{eq:exp_p4}
    \braket{\hat{p}^4(t)} = \left.  \frac{\partial^4 }{\partial x^4}  \chi (\bar{r}, t) \right|_{\bar{r}=0}  = \frac{3}{4} \left(\Delta_{p^2}\right)^2 + 3  \Delta_{p^2}  p^2 + p^4 
\end{equation}
\begin{equation}
\label{eq:exp_xp}
    \braket{\hat{x}^2 \hat{p}^2 + \hat{p}^2 \hat{x}^2} = \left. \left( \frac{\partial^2 }{\partial x^2} \frac{\partial^2 }{\partial p^2}  + \frac{\partial^2 }{\partial p^2} \frac{\partial^2 }{\partial x^2} \right)  \chi (\bar{r}, t) \right|_{\bar{r}=0}  = 4 \left(\Delta_{\{xp\}}\right)^2 + 4 \Delta_{\{xp\}} x p +  \left( \Delta_{p^2} + 2 p^2 \right) \left(\frac{1}{2}    \Delta_{x^2} + x^2 \right) 
\end{equation}
where we dropped the time dependence in the first and second moments. 

For the discussion in Sec.~\ref{app:decoherence} it is useful to compute the characteristic functions of each element of the spin density at the end of the interferometry. The last displacement only changes the first moments, and from Eq.~\ref{eq:characterstic_time}, the diagonal terms are
\begin{equation}
\label{eq:characterstic_diagonal}
     \chi_{1,1}(\bar{r},t_f) =\chi_{-1,-1}(\bar{r},t_f) = e^{- \frac{1}{4} \bar{r}^T \Omega^T S^{-1}_\text{tot}(t)  \left(S^{-1}_\text{tot}(t) \right)^T \Omega \bar{r} \pm i \bar{r}^T \Omega^T  \left( S^{-1}_\text{tot}(t)r_0 + r_0 \right)}
\end{equation}
The off-diagonal terms can be computed accordingly
\begin{align}
\label{eq:characterstic_off_diagonal}
    \chi_{1,-1}(\bar{r},t) &=  \text{Tr} \left[ \ket{0}\bra{0} \hat{\mathcal{D}}_{r_0}  \hat{S}_\text{tot}^\dag \hat{\mathcal{D}}_{r_0} \hat{\mathcal{D}}_{- \bar{r}} \hat{\mathcal{D}}_{ r_0}  \hat{S}_\text{tot} \hat{\mathcal{D}}_{ r_0}   \right] = \text{Tr} \left[ \ket{0}\bra{0} \hat{S}_\text{tot}^\dag \hat{\mathcal{D}}_{S^{-1}_\text{tot} r_0} \hat{\mathcal{D}}_{2 r_0- \bar{r}} \hat{\mathcal{D}}_{S^{-1}_\text{tot} r_0}\hat{S}_\text{tot}    \right]
     \\ 
    &= \text{Tr} \left[ \ket{0}\bra{0} \hat{S}_\text{tot}^\dag \hat{\mathcal{D}}_{2 \left( S^{-1}_\text{tot} + \mathds{1} \right) - \bar{r}} \hat{S}_\text{tot}    \right] = \text{Tr} \left[ \ket{0}\bra{0} \hat{\mathcal{D}}_{S^{-1}_\text{tot} \left[ 2 \left( S^{-1}_\text{tot} + \mathds{1} \right) r_0 - \bar{r} \right]}   \right] \\
    &= e^{- \frac{1}{4} \left[ \bar{r} - 2\left( r_0 + S^{-1}_{\text{tot}} r_0 \right) \right]^T \Omega^T \sigma(t) \Omega \left[ \bar{r} - 2\left( r_0 + S^{-1}_{\text{tot}} r_0 \right) \right]}
\end{align}
which is equal to $\chi_{-1,1}(\bar{r},t)$ as it is real. Thus, while the off-diagonal terms are non-Gaussian (for instance, in their Wigner function), their characteristic function is still fully determined by the first and second moments \cite{serafini_minimum_2004}.

\section{Decoherence Analysis \label{app:decoherence}}

The following analysis applies to both the diagonal terms of the density matrix -- which are Gaussian -- and to the off-diagonal ones -- as their characteristic functions are also Gaussian \cite{serafini_minimum_2004}. In fact, Eq.~\ref{eq:characterstic_diagonal} and Eq.~\ref{eq:characterstic_off_diagonal} are Gaussians fully characterized by the first and second moments. Any Markovian master equation for continuous variables with environmental coupling of second order in $\hat{x}$ and $\hat{p}$ (such as the one in Eq.~\ref{eq:master_pos}) can be recast to the form:
\begin{equation}
\label{eq:general_master}
    \frac{\partial \rho}{\partial t} =\sum_{j,k} \left[\left( \frac{i}{2} \Omega A  -\frac{\Omega D \Omega^T}{4} \right)_{j,k} \hat{r}_j \hat{r}_k \rho + h.c. \right]  + \left[\left( \frac{i}{2} \Omega A + \frac{\Omega D \Omega^T}{4} \right)_{j,k}\hat{r}_j  \rho \hat{r}_k + h.c. \right] \;, 
\end{equation}
where $A$ and $C$ are known as the drift and diffusion matrices, respectively \cite{serafini_quantum_2017}. The dynamics is fully determined in terms of the time evolution of the first and second moments. Eq.~\ref{eq:general_master} implies 
\begin{equation}
    \frac{\partial r}{\partial t} = A r + d \;,\hspace{1 cm} \frac{\partial \sigma}{\partial t} = A \sigma + \sigma A^T + D \;,
\end{equation}
where $d$ is an additional driving term and the latter equation is known as the Lyapunov equation. They have general solutions
\begin{equation}
\label{eq:first_swecond_master}
    r(t) = e^{A t} \left( r(0) + d \right)  \; , \hspace{1cm} \sigma(t) = \int_{0}^t \text{d} t' e^{A(t-t')}D e^{A^T(t-t')} + e^{A t}  \sigma(0) e^{A^T t} \;.
\end{equation}
By comparing Eq.~\ref{eq:master_pos} and Eq.~\ref{eq:general_master}, it follows that
\begin{equation}
    A_\pm =  \Omega H_\pm = \begin{pmatrix}
        0 & 1 \\ 
        \mp 1 & 0
    \end{pmatrix}\;, \hspace{1cm}
    D = \begin{pmatrix}
        0 & 0 \\ 
        0 & \Gamma_x
    \end{pmatrix}\;.
\end{equation}
Hence,  $e^{A_\pm t}  = S_{H_\pm}^{-1}$. If there is no external force, $d = 0$. This implies that the first moments are unchanged under this source of error. At the end of the interferometer, they are still given by $r_\pm(t_f) = \pm S_{\text{tot}} r_0 \pm  r_0 = 0$. 
Hence, the density matrix of the qubit will be unchanged. This trivially follows by taking the trace of the spatial degree of freedom using the characteristic functions in  Eq.~\ref{eq:characterstic_diagonal} and Eq.~\ref{eq:characterstic_off_diagonal} with $\bar{r} = 0$ (and always equal to $1$ for any evolution of $\sigma (t)$). 

For completeness, we include the second moment. The following integrals are useful
\begin{equation}
\label{eq:D_plus}
    D_{t_+} =  \int_{0}^{t_+} \text{d} t' e^{A_+(t_+ -t')}D e^{A_+^T(t_+ -t')} =  \frac{1   }{4}\Gamma_x
\begin{pmatrix}
     \pi & 2\\
  2& \pi
\end{pmatrix}
\end{equation}
\begin{equation}
\label{eq:D_minus}
    D_{t_-} =  \int_{0}^{t_-} \text{d} t' e^{A_-(t_- -t')}D e^{A_-^T(t_- -t')} =  \frac{\Gamma_x}{4} 
\begin{pmatrix}
 \sinh (2 t_-)- 2 t_- & 2 \sinh ^2 t_- \\
2 \sinh ^2 t_- & \sinh (2t_-) + t_- \\
\end{pmatrix}
\end{equation}
where the last equality of Eq.~\ref{eq:D_plus} is for $t_+ = \pi/2$. By repetitively using Eq.~\ref{eq:first_swecond_master}, the covariance matrix at the end of the interferometer is given by
\begin{align}
    \sigma_\Gamma(t_f) & =  D_{t_-}+  S_{H_-}  D_{t_+} S_{H_-}^T + S_{H_-} S_{H_+}  D_{t_-}  S_{H_+}^T S_{H_-}^T +   S_{H_-}S_{H_+} S_{H_-}  D_{t_+} S_{H_-}^T S_{H_+}^TS_{H_-}^T  +    S_{\text{tot}}\sigma(0) S_{\text{tot}}^T 
\end{align}
The first four terms of the expression describe diffusion during the protocol and the last the protocol itself. However, they do not affect spin measurements. 

To include the spin decoherence it is possible to solve the Master equation given in Eq.~\ref{eq:master_spin}. The protocol Hamiltonians commute with both the spin-free evolution as well as the Lambladian term. Then, the off-diagonal term of the single qubit density matrix at the end of the protocol decays as $e^{- \Gamma_q/\omega t_f}$. Similarly, in the case of two qubits (with the entangling dynamics), this can be compactly written as
\begin{equation*}
   \tilde{\rho}_{j,k}^{m,n} = \rho_{j,k}^{m,n} e^{-(2-\delta_{j m}-\delta_{k n})\Gamma_q/\omega t_f} \;,
\end{equation*}
where, $\tilde{\rho}_{j,k}^{m,n}$ are the decohered density matrix elements and $\rho_{j,k}^{m,n}$ are given in Eq.~\ref{eq:density_general_2}. The PPT criterion under this source of decoherence,  is plot in following Figure. 

\begin{figure}[!]
\includegraphics[width=0.9\textwidth]{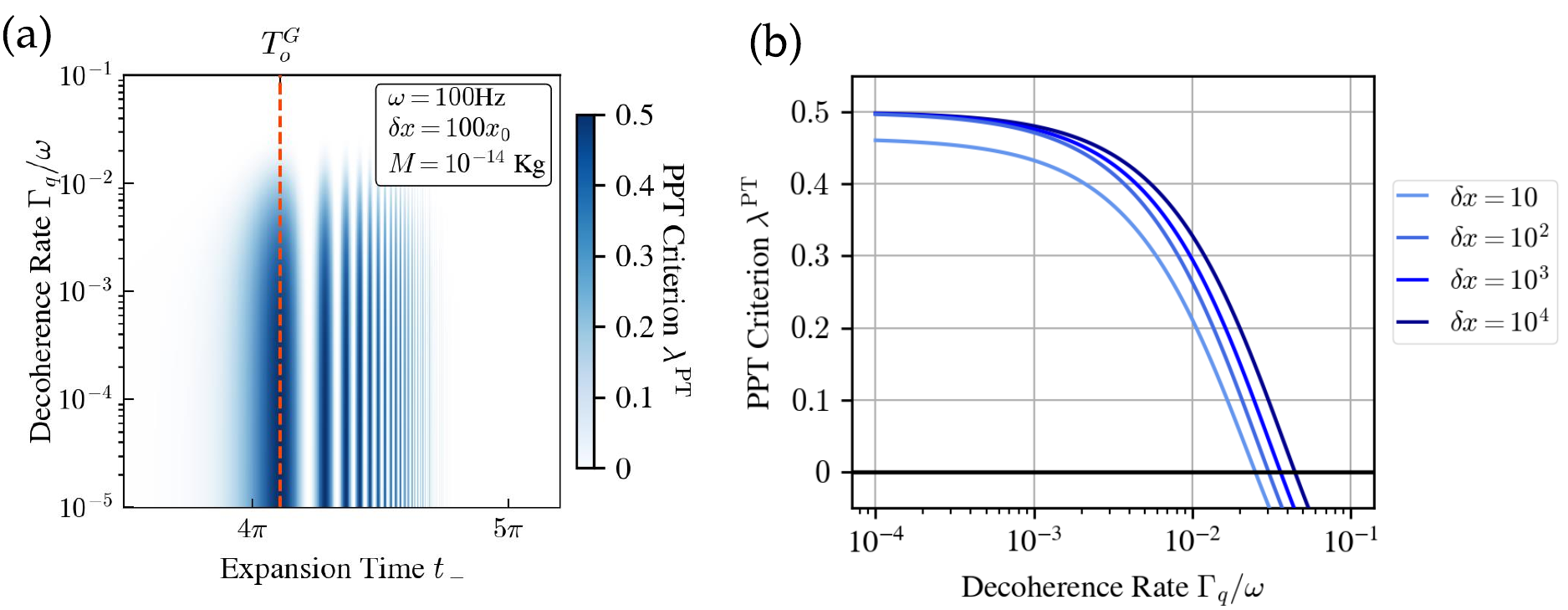}
  \caption{\label{fig:dec} PPT criterion under qubit decoherence: (a) entanglement as function of decoherence rate $\Gamma_q$ and expansion time $t_-$; (b) Entanglmenet at optimal time $T_o^G$ as function of $\Gamma_q$  for different initial superposition sizes $\delta x$.}
\end{figure}

\section{Sudden Approximation \label{sm:linearity_switching}}

Given the unitarity of the protocol, the sudden approximation allows us to express it as in Eq.~\ref{eq:unitary_ideal}. Let us define the time-dependent switching Hamiltonian and its average as
\begin{equation}
    \hat{H}(t) = \frac{\hbar \omega }{2} \left( f(t) \hat{x}^{2} + \hat{p}^2 \right) \; ,\hspace{1cm}
    \hat{H}_A = \frac{\hbar \omega }{2} \left( \braket{f} \hat{x}^{2} + \hat{p}^2 \right)\;,
\end{equation} 
where $t\in  [0,1]  $, $f(0) = - 1$, $f(1) = 1$ and $ \braket{f}$ is the average of the function over $t$. This implies that $\hat{H}(0) = \hat{H}_-$ and $\hat{H}(1) = \hat{H}_+$. The adiabatic approximation holds if $ \delta t \ll \hbar ( \Delta_{ H_A })^{-1}$ where $\Delta_{ H_A }= \sqrt{\braket{\hat{H}_{A}^2} - \braket{\hat{ H }_A }^2 }$ and $\delta t$ is the switching time interval. Then, 
\begin{equation}
    \Delta_{H_A}^2 = \frac{1}{2} \left( \Delta_{p^2}^2+4  \Delta_{p^2} p^2+\braket{f}\left( \braket{f} \Delta_{x^2}  \left( \Delta_{x^2}+4 x^2\right)+2  \Delta_{\{x p \}}^2+8  \Delta_{\{x p \}} p x \right)\right) \;,
\end{equation}
where Eq.~\ref{eq:exp_x3x4},~\ref{eq:exp_p4} and~\ref{eq:exp_xp} are used. We focus the discussion at the time of the protocol $t_s = t_+ + t_-$ as the mass has its the switching point when the first and second moments are mostly delocalized. It follows that  
\begin{align}
    \Delta_{H_A}^2(t_s) &=  \frac{1}{4} \left(2 \delta x^2+1\right) \left[\left(\braket{f}+1 \right)^2 \cosh (4 t_-)+\left(\braket{f}-1 \right)^2\right]+ \left(1 - \braket{f}^2\right) \delta x^2 \cosh (2 t_-) \\
    &\approx \frac{1}{8} \left(2 \delta x^2+1\right) \left(\braket{f}+1 \right)^2 e^{4 t_-}
\end{align}
where the second line is the leading term in $t_-$. By considering $\braket{f} \approx 0$, \textit{i.e.} the switching function is approximately symmetric, Eq.~\ref{eq:sudden_approx} follows.



\section{Humpty-Dumpty Effect}

The Humpty-Dumpty effect happens when there is not a perfect recombination of the two branches of the interferometer. This causes a loss of visibility in the spin measurement as the off-diagonal term of the density matrix in Eq.~\ref{eq:density_general_1qubit} decays exponentially if $S_{\text{tot}} \neq - \mathds{1}$.

\begin{figure}[!b]
\includegraphics[width=0.4\textwidth]{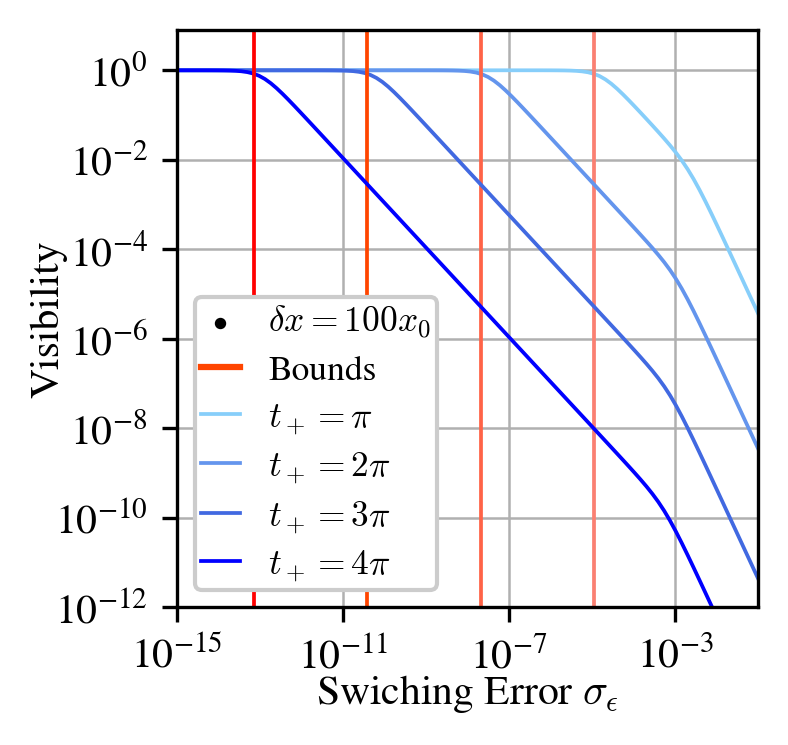}
  \caption{\label{fig:humpy} Humpty-Dumpty effect as a function of the standard deviation of the switching time with error ($\sigma_\epsilon$) for different expansion times.}
\end{figure}

The potentials of the protocol have to alternate according to the switching times  $t_i\in \{\frac{\pi}{2}, t_-,\frac{\pi}{2}, t_-\}$. Hence, there are four possible errors $t_i^\epsilon = t_i\left( 1 + \epsilon_i \right)$, where $\epsilon_i$ are random variables with probability distribution $P(\epsilon_i) = \mathcal{N}(0,\sigma_\epsilon^2)$, where $\mathcal{N}$ is a Gaussian distribution with standard deviation $\sigma_\epsilon$ (for simplicity taken to be identical among distributions). Only three are independent as it is always possible to scale $t_- \to t_-  - \epsilon_2 $  and define $\epsilon_{4} \to \epsilon_{24} = \epsilon_2 -  \epsilon_4$, having $P(\epsilon_{24}) = \mathcal{N}(0,2 \sigma_\epsilon^2)$. Let us now define the symplectic transformation with errors $S_\epsilon$, given by $\epsilon = (\epsilon_1, \epsilon_3, \epsilon_{24})$ and the corresponding density matrix at the end of the protocol $\rho_\epsilon$. Then, with the number of runs of the experiment tending to infinity, the total density matrix is given by:
\begin{equation}
    \rho^{\text{HD}} = \int \text{d} \epsilon_1 \text{d} \epsilon_{24} \text{d} \epsilon_3 P(\epsilon_1) P(\epsilon_{24})P(\epsilon_{3}) \rho_\epsilon 
\end{equation}
We study the case where no external force is applied ($r_{\text{tot}} = 0$). Then, the first and second moments with errors are $r_\epsilon(t_f) = S_\epsilon^{-1} r_0$ and $\sigma_\epsilon(t_f) = S_\epsilon^{-1} \sigma_0  \left( S_\epsilon^{-1} \right)^T$.  From Eq.~\ref{eq:density_general_1qubit}, the only non-trivial contribution is given for the off-diagonal term, according to
\begin{align}
\label{eq:humpy_dumpy_OG}
    \rho^{\text{HD}}_{0,1} = \rho^{\text{HD}}_{1,0} &=  \int \text{d} \epsilon_1 \text{d} \epsilon_{24} \text{d} \epsilon_3 \mathcal{N}(0, \sigma_\epsilon^2) \mathcal{N}(0, \sigma_\epsilon^2) \mathcal{N}(0,2 \sigma_\epsilon^2)
    e^{ -   \left(r_0 + r_\epsilon(t_f) \right)^T \Omega^T \sigma_\epsilon(t_f) \Omega \left(r_0 + r_\epsilon(t_f) \right)} \\ 
    &\approx \frac{4 \sqrt{2}}{\sqrt{\delta x^2 \sigma_{\epsilon} ^2 \left(16 t_-^2+\pi ^2\right) \left(8-\pi ^2 \delta x^2 \sigma_{\epsilon} ^2\right)+\pi ^2 \delta x^2 \sigma_{\epsilon} ^2 \cosh (4 t_-) \left(\delta x^2 \sigma_{\epsilon} ^2 \left(16 t_-^2+\pi ^2\right)+8\right)+32}}
\end{align}
where the second equality is followed by Taylor expanding the exponent up to the second order in the $\epsilon$s (assumed to be small compared to 1). For large $t_-$, the leading term is given by
\begin{equation}
\label{eq:humpy_dumpy}
    \rho^{\text{HD}}_{0,1} \approx \frac{8}{\pi \delta x \sigma_\epsilon} \frac{e^{-2 t_-}}{\sqrt{8 + \delta x^2 \sigma_\epsilon^2 \left( \pi + 16 t_- \right)}}
\end{equation}
In order to have visibility of order $\mathcal{O}(1)$, Eq.~\ref{eq:humpy_dumpy} can be compared to $1$ and inverted to find the bound on $\sigma_\epsilon$, which leads to Eq.~\ref{eq:bound_humpy_dumpy}. The visibility as a function of $\sigma_\epsilon$ is given in Fig. \ref{fig:humpy}

\end{document}